\documentclass[aps,prx,groupedaddress,reprint,showpacs,pdftex]{revtex4-1}

\usepackage{graphicx}
\usepackage{tabularx}
\usepackage{dcolumn}
\usepackage{bm}
\usepackage{amsmath}
\usepackage{amssymb}
\usepackage{txfonts}
\usepackage{url}
\usepackage[usenames,dvipsnames]{color}
\usepackage{comment}
\usepackage{here} 

\definecolor{Gray}{gray}{0.0}
\definecolor{lightGray}{gray}{0.35}
\begin{document}
\title{
Stochastic estimations of a total 
number of classes for the clusterings 
with too enormous samples to be accommodate into 
a clustering engine 
}
\author{Keishu Utimula$^{1}$}
\email{mwkumk1702@icloud.com}
\author{Genki I. Prayogo$^{1}$}
\author{Kousuke Nakano$^{2}$}
\author{Kenta Hongo$^{3,4,5}$}
\author{Ryo Maezono$^{2}$}

\affiliation{$^{1}$
  School of Materials Science, JAIST, Asahidai 1-1, Nomi, Ishikawa,
  923-1292, Japan
}

\affiliation{$^{2}$
  School of Information Science, JAIST, Asahidai 1-1, Nomi, Ishikawa,
  923-1292, Japan
}  


\affiliation{$^{3}$
  Research Center for Advanced Computing Infrastructure,
  JAIST, Asahidai 1-1, Nomi, Ishikawa 923-1292, Japan
}
\affiliation{$^{4}$
  Center for Materials Research by Information Integration,
  Research and Services Division of Materials Data and
  Integrated System, National Institute for Materials Science,
  Tsukuba 305-0047, Japan
}
\affiliation{$^{5}$
  PRESTO, Japan Science and Technology Agency, 4-1-8 Honcho,
  Kawaguchi-shi, Saitama 322-0012, Japan
}
\date{\today}
\begin{abstract}
We considered the problem how to 
handle the exploding number of 
possibilities to be sorted into 
irreducible classes by using 
a clustering tool when 
its input capacity cannot accommodate 
the total number of the possibility. 
Concrete situations are explained 
taking examples of atomic substitutions 
in the supercell modeling of alloys. 
The number of the possibility 
sometimes amounts to $\sim$ trillion 
being too large to be accommodate. 
It is hence not practically 
feasible to identify 
how many irreducible classes exist 
by straightforward manner 
even though there are several tools 
available to perform the clustering. 
We have developed a stochastic framework 
to avoid the shortage of capacity, 
providing a method to estimate 
the total number of irreducible classes 
(the order of the classes) 
as a statistical estimate. 
A prominent conclusion derived here is 
that the statistical variation of 
the number of classes at each 
sampling trial is working 
as a promising measure to estimate 
the order. 
\end{abstract}
\maketitle

\section{Introduction}
\label{sec.intro}
It is an omnipresent task 
to classify a large number of data 
into some groups by individual attributes. 
Beyond the handling based on clear 
classification rules, 
the clustering driven by 
unsupervised learning techniques 
has well been developed
~\cite{2014KUE, 2017SUR, 2017IWA, 2018STA}
%
An innately concomitant situation 
along the purpose of the clustering is
the large amount of data to be classified.
The larger it gets, 
the more accurate in general 
the statistical quality of the estimations 
becomes, showing the superiority 
of data science. 
However, the size of data 
can often be too large to be 
accommodated into a single clustering 
process. 
Let us set up the problem as follows: 
Suppose we have a clustering tool 
that can process up to $l_{\rm max}$ samples. 
The tool classifies $l$ reducible samples 
(as input) 
into $M(l)$ irreducible groups (as output) 
based on some attributes. 
Let the maximum possible number of the samples 
be $L~(\ge l_{\rm max})$, 
$G=M(L)$ is the total number of 
the irreducible attributes that we want to know. 
However, we cannot identify $G$ 
when $l_{\rm max} \ll L$ because of 
the limitation of the input capacity. 

\vspace{2mm}
To illustrate the problem more concretely, 
let us consider a magnetic alloy, 
Nd$_{(1-x)}$Ce$_{x-y}$La$_{y}$)$_2$Fe$_{14}$B, 
which is composed based on Nd$_2$Fe$_{14}$B 
by substituting a part of Nd-sites 
with Ce or La. 
Supercell methods in {\it ab initio} calculations 
are used to describe such atomic substitutions 
modelled as a periodic array of a cell structure.
~\cite{2020UTI}
%
The size of the cell (specified by 
how many Nd to be accommodated) is 
chosen so that the given concentration $(x,y)$ 
is captured by the ratio of the 
number of atoms.
~\cite{2020UTI}
%
To describe 
Nd$_{0.7}$Ce$_{0.225}$La$_{0.075}$)$_2$Fe$_{14}$B, 
40 Nd-sites are substituted 
by nine Ce and three La. 
The number of possible
atomic configurations in this case
would amount to
40!/(28!9!3!) = 1,229,107,765,600 = $L$ 
(total reducible samples to be classified). 
These 'raw' configurations
can further be classified into
subgroups of {\it irreducible} configurations 
based on the theoretical framework to
identify equivalent structures 
under the symmetric operations.
~\cite{2007PAU}
%
The number of the subgroup, $G$, 
gets reduced by several orders, 
typically $G\sim 100$ 
even for $L\sim$ trillion. 
%
Once we identify $G$, 
we can construct concrete $G$ 
representative configurations 
as irreducible structures 
by a recursive way 
%
Several packages in Materials Science 
provide such a tool 
to perform the classifications
~\cite{2005STO, 2018MAT}
but their input capacity, 
$l_{\rm max}$, cannot handle 
such a large number of $L\sim$ 
trillion. 

\vspace{2mm}
The present study has worked on 
the problem to identify $G$ even 
for $l_{\rm max} \ll L$. 
A stochastic method is proposed 
to identify $G=M(L)$ utilizing a tool 
giving $M(l)$~($l\ll L$) 
with $l_{\rm max} \ll L$. 
The method repeats the numbers of 
the sampling to calculate $M(l)$, 
getting the statistics, 
$\left\{ M(l) \right\}$, 
by which it estimates $G$. 
The estimation can be achieved 
in far fewer trials than $L/l$, 
as shown in this study. 
For the above example 
of the classification of 
crystal structures, 
we note that there is more powerful, 
deterministic framework 
(Polya's theorem), 
though the present method can 
handle it as well. 
However, the method described 
in this paper can handle 
a broader category of the problem. 
The method only relies on  
a tool to provide $M(l)$, 
which can be a machine-learning 
tool to perform a clustering, 
where the classification rule 
is not necessary to be based on 
any clear principles like 
solid state physics, group theory 
{\it etc.} as in the example. 
When combined with 
existing clustering tools, 
the method can expand 
the scope of the tools 
beyond the limitation 
of input capacities 
(depending on the memory capacity 
in the computations).

\begin{figure}[htbp]
  \includegraphics[width=\hsize]{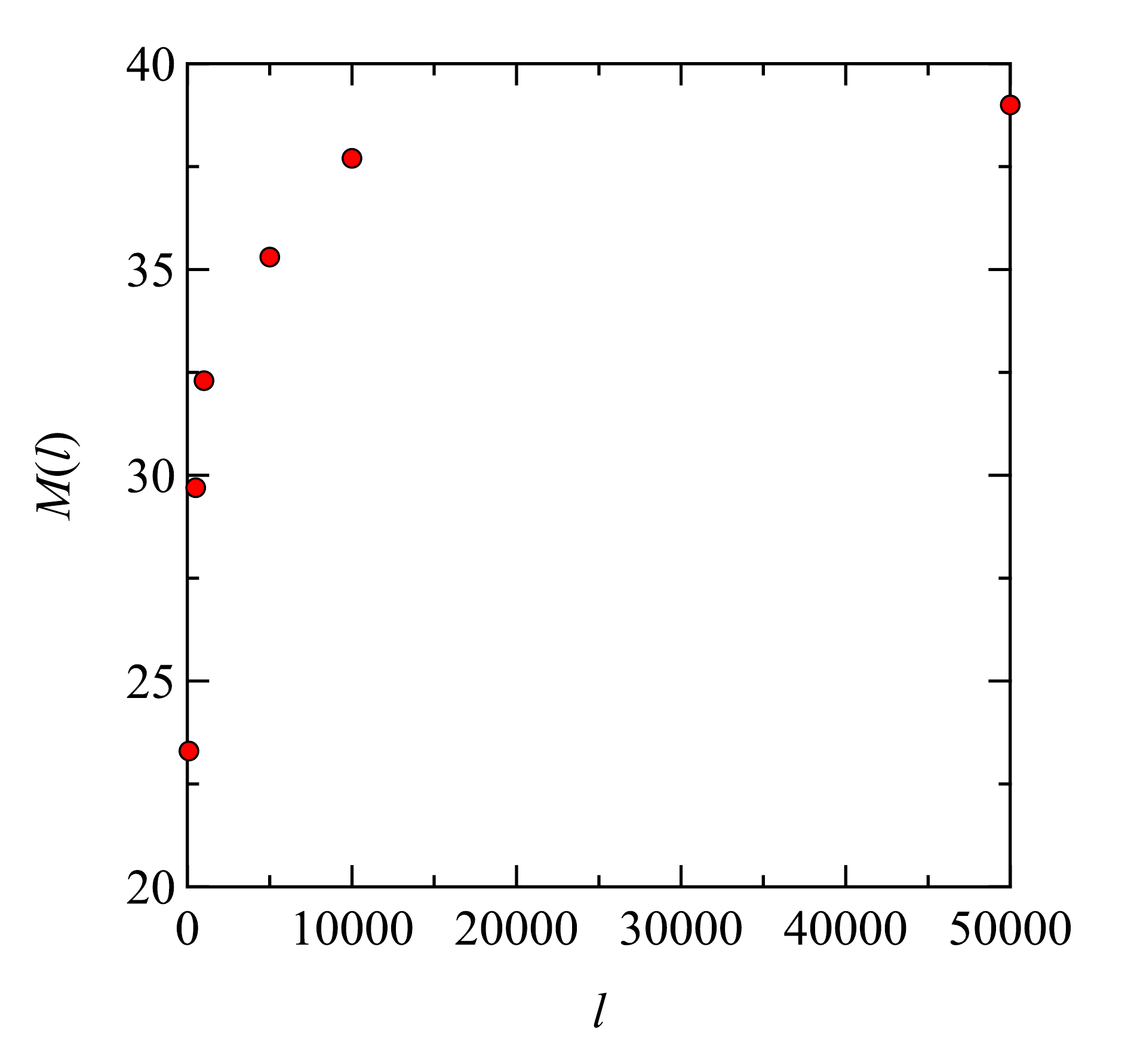}
  \caption{
    Estimation of 
    the number of irreducible classifications 
    ($M$) for given $l$ reducible samples. 
    As $l$ increases to some extent, 
    the estimated $M(l)$ gets converged into 
    the true number $M(L)$ ($L$ is the maximum
    possible size of reducible samples). 
    The plot is taken from the case of
    Nd$_{0.7}$Ce$_{0.225}$La$_{0.075}$)$_2$Fe$_{14}$B.    
  }
  \label{substSample}
\end{figure}

\vspace{2mm}
If $l_{\rm max}$
can get around several times of $G$,
we will see the convergence of $M(l)\to G$
to identify the number of irreducible groups $G$,
as shown in Fig.~\ref{substSample}.
This strategy still requires costs because, in general,
it is laborious to identify whether a monotonic behavior
has reached at the convergence, still requiring
enough number of $l$.
The method explained here
gives a powerful breakthrough,
providing another way to identify $G$ as a position
of the peak of a measure.
We derived that $V[M(l)]$ (the variance of $M(l)$)
with respect to the random choice of $l$ members
of reducible samples 
takes a maximum when $l\sim G$.

\section{Outline of the results}
To formulate the variance of $M(l)$,
we shall introduce a probability for a set of 
$l$ samples to be sorted into
$M$ irreducible classes, denoted as 
$P\left(l,M;G\right)$.
The expectation value,
$\bar M$, and its variance $V[M(l)]$ are then given as 
\begin{eqnarray}
  \bar M\left( l \right)
  &= \sum\limits_{M = 1}^{G} {M \cdot P\left( {l,M;G} \right)}
  \ ,
  \\
  V\left[M(l) \right]
  &= \sum\limits_{M = 1}^{G} {
    {{\left( {M - \bar M \left( l \right)} \right)}^2}
   \cdot P\left( {l,M;G} \right) 
  } \ .
  \label{aveAndvar}
\end{eqnarray}
For $l=1$, the sample surely
corresponds to a irreducible class,
namely $\bar M(0)=1$
regardless of the choice of the
sample;
hence $V[M(1)]=0$.
In another limit with larger $l$,
we know the convergence $M(l\to L)=G$,
and hence $V[M(l\to L)]\to 0$.
As such, we expect $V[M(l)]$ to have
a maximum peak as a function of $l$
in the range of $1<l<L$.
We can actually derive that the
$V[M(l)]$ has
a peak at 
around $l\sim G$,
shown as follows.  
\begin{figure}[htbp]
  \includegraphics[width=\hsize]{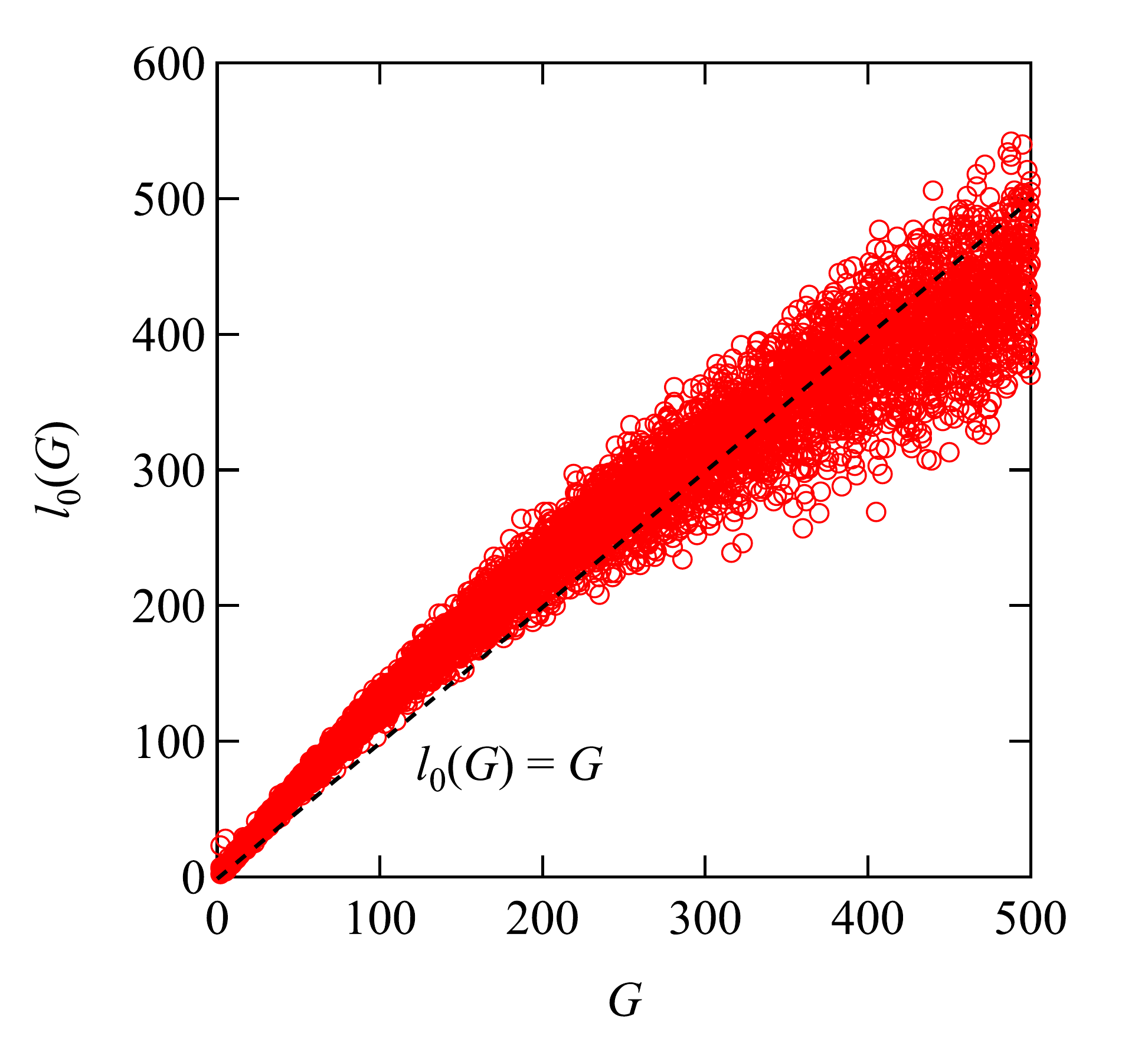}
  \caption{
    Plots of $l_0$ at which the variance $V[M(l)]$ has a peak.
    The results are obtained by numerical simulations
    with several different sizes 
    $G$.
    For each $G$, we generated ten different cases of
    the multiplicities for each irreducible structure randomly.
    Then we naively picked up a sampling with $l$ to identify
    $M(l)$, and then evaluate the variance for each $l$
    to get its behavior with a peak at $l=l_0$.
    The plot is observed to be scaling with $l_0(G)=G$
    broken black 
    line),
    leading to a conclusion that we can estimate $G$ from
    the position of the peak of $V[M(l)]$. 
  }
\label{maxVwithoutApprox}
\end{figure}

\vspace{2mm}
With the most simplified assumption
that the size of each
irreducible class
({\it i.e.}, the number of
elements inside each class)
is identical, 
we can derive an asymptotic behavior, 
\begin{eqnarray}
  P\left( {l,G,M} \right)
  & \sim&
  \left\{ {1 - C\exp \left[ { - \left( {\frac{{G - M + 1}}{G}} \right)
        \left( {l - M} \right)} \right]} \right\}\nonumber\\
  && \quad \times \exp \left[ { - \left( {\frac{{G - M}}
        {G}\left( {l - M} \right)}
      \right)} \right] \ ,
  \nonumber \\
    C &=& \left( {1 - \cfrac{{_G{C_M} \cdot M!}}{{{G^M}}}} \right) \ , 
  \label{prob_norm}
\end{eqnarray}
as given in the appendix, 
\S\ref{der.prob}$\sim$
\S\ref{asymptotic}, 
and shown in Fig.~\ref{P_approx}.
By substituting this into Eq.(\ref{aveAndvar}),
we can show that the variance $V\left[M(l) \right]$
has a peak around $l\sim (G+1)$ (\S\ref{max.var}).
Even without such an assumption used in the analytical treatments,
numerical verifications have shown that the conclusion
is kept unchanged, 
as shown in Fig.~\ref{maxVwithoutApprox}.
\begin{figure}[htbp]
  \includegraphics[width=\hsize]{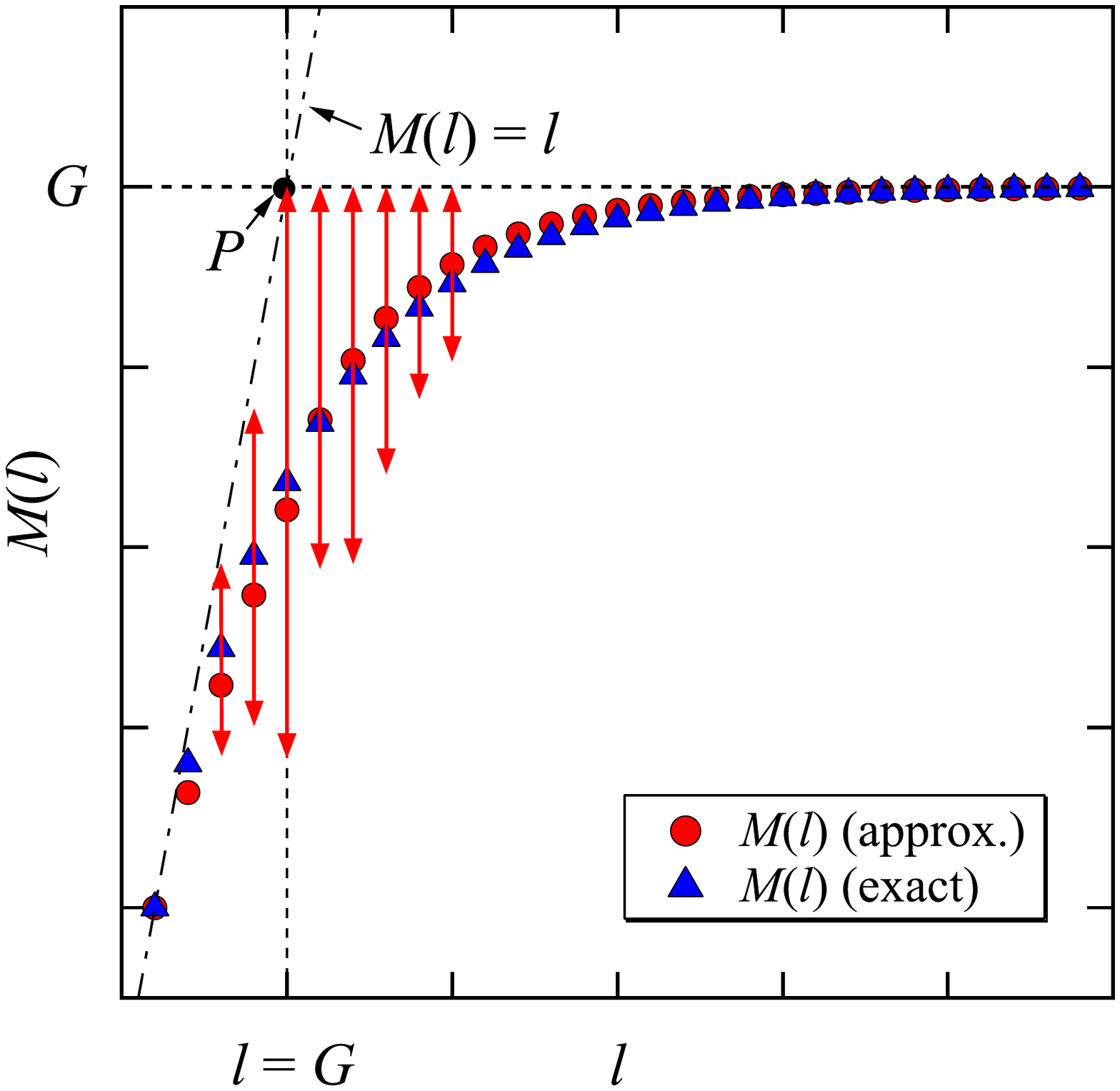}
  \caption{
    A schematic picture to show the variance of $M(l)$
    would take a maximum at $l\sim G$. At fewer (larger) $l$,
    $M(l)$ is expected to behave like $M(l)=l$ ($M(l)=G$).
    Exact solution (red line) and approximated solution
    (blue line) of $M(l)$.
    These are derived from equation (\ref{prob_exact})
    and (\ref{prob_norm}), respectively.
    Vertical red arrows schematically show the widths of
    variances around it.
    Since the maximum possible value for $M(l)$ is $G$,
    the tail of the variance cannot go beyond $M(l)=G$,
    and then the most widest possible tail would occur
    at the intersection $P$ in the figure, identified as $l=G$. 
  }
  \label{teisei}
\end{figure}

\vspace{2mm}
We can get a rough understanding that the $V[M(l)]$
has a peak at 
around $l\sim G$ from a
schematic picture as shown in Fig.~\ref{teisei}.
For tiny sizes of samplings, $l\sim 1,2,\cdots$,
we expect $M(l)= l$ with little duplications,
specifying the raising up of the behavior at smaller $l$.
The convergence, $M(l)\to G$, at larger $l$, 
then leads to the schematic behavior, 
as shown in Fig.~\ref{teisei}.
By vertical red arrows on the dependence $M(l)$,
we schematically show the widths of variances around it.
Since the maximum possible value for $M(l)$ is $G$,
the tail of the variance cannot go beyond $M(l)=G$,
and then the widest 
possible tail
would occur at the intersection $P$ in the figure,
identified as $l=G$. 

\section{Discussions}
\begin{figure}[H]
  \includegraphics[width=\hsize]{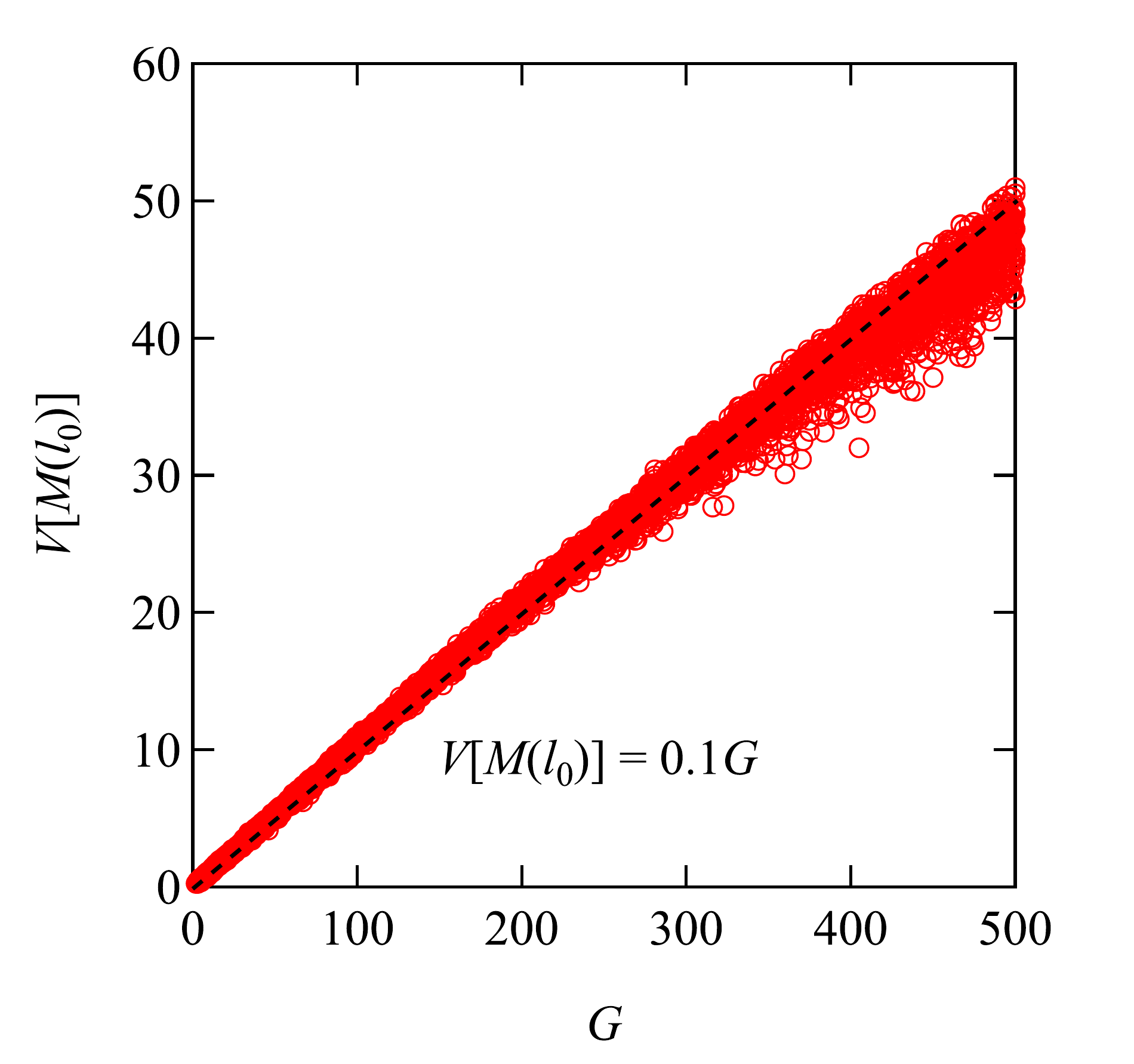}
  \caption{
  Plots of the maximum variance, $V[M(l_0)]$, 
  for difference $G$. 
  For each $G$, we generated ten different cases 
  of the multiplicities for each irreducible 
  structure randomly. 
  Then we naively picked up a sampling with $l$ 
  to identify $M(l)$, and then evaluate 
  the variance for each $l$ to get its behavior 
  with a peak at $l=l_0$. 
  The plot is observed to be scaling with 
  $V=0.1\times G$(blue broken line), 
  leading to a conclusion that we can estimate 
  $G$ from the value with $V[M(l_0)]$ multiplied by ten. 
  }
  \label{heuV}
\end{figure}

Though that's still within the extent 
of a heuristic finding, we note that 
$10\times V[M(l_0)]$ provides a quite 
reliable estimate for $G$
($l_0$ denotes the peak position
of $V[M(l_0)]$). 
As shown in Fig.~\ref{heuV}, the estimate 
seems rather firm than another estimator, 
'$l_0 \sim G$ in Fig.~\ref{maxVwithoutApprox}. 

\section{Conclusion}
\label{sec.conc}
We have formulated 
a stochastic framework
to handle the vast number of 
samples ($L \sim$ trillion)
to be classified
by any clustering tool
to find out the total
number of irreducible classes $G$. 
It is sometimes the situation
we considered
that $L$ is too large to be
accommodated into the
capacity of the input size of
the tool because of the limitation
of the memory- or file-size. 
Among $L$, the framework randomly picks up
a set of samples with the size $l$ ($\ll L$)
being affordable by the capacity,
to be sorted by a clustering tool
into $M$ classes.
By repeating the sampling
with varying the size $l$, 
we get the statistics
$\left\{M(l)\right\}$.
We have derived that
the variance of $M(l)$
takes maximum at
$l\sim G$. 
Though there is no rigid mathematical 
verifications, we heuristically found
that $G$ is 
estimated as the quantity around 
ten times of $l_0$. 

\section{Acknowledgments}
The computations in this work have been performed 
using the facilities of 
Research Center for Advanced Computing 
Infrastructure at JAIST. 
R.M. is grateful for financial supports from 
MEXT-KAKENHI (19H04692 and 16KK0097), 
from FLAGSHIP2020 (project nos. hp190169 and hp190167 at K-computer), 
from Toyota Motor Corporation, from I-O DATA Foundation, 
from the Air Force Office of Scientific Research 
(AFOSR-AOARD/FA2386-17-1-4049;FA2386-19-1-4015), 
and from JSPS Bilateral Joint Projects (with India DST). 
K.H. is grateful for financial supports from 
FLAGSHIP2020
(project nos. hp190169 and hp190167 at K-computer),
KAKENHI grant (17K17762 and 19K05029),
a Grant-in-Aid for Scientific Research on Innovative Areas (16H06439 and 19H05169), 
PRESTO (JPMJPR16NA) and
the ``Materials research by Information Integration Initiative" (MI$^2$I) 
project of the Support Program for Starting Up Innovation Hub 
from Japan Science and Technology Agency (JST).
K.U. is grateful for financial supports from JAIST Research Grant (Fundamental Research) 2019. 


\newpage
\section{Appendix}
\label{sec.a}

\subsection{Derivation of the probability}
\label{der.prob}
We denote the probability 
to get $M$ irreducible classes 
after a clustering tool
processes over $l$ reducible samples 
as $P\left( {l,M;G} \right)$, 
where $l$ samples are
randomly taken from 
a population having total
$G$ irreducible classes. 
The situation is equivalently
described like 
'distributing color envelops  
into $l$ pigeonholes with one each, 
where an envelop has 
$G$-color variation,
but only $M$ ($<G$)
colors are found over
the pigeonholes'. 
Though the multiplicity
for each color
({\i.e.}, the number of
envelops with each color) 
is generally different each other,
we put an assumption that the
multiplicity is the same for
each color.
We can then evaluate $P\left( {l,M;G} \right)$ 
in more simplified manner analytically,
though we shall examine later whether
the final conclusion is not so affected
by the assumption. 
Letting the number of cases be 
$a\left( {l,M;G} \right)$, 
the probability is written as 
\begin{eqnarray}
  P\left( {l,M;G} \right)
  &=& a(l,M;G)/\sum_{i=1}^{G}{a(l,i;G)} \ .
\label{p_first}
\end{eqnarray}
It can be denoted that 
\[
a\left( {l,M;G} \right) = {}_G{C_M}\times F \ , 
\]
where $F\sim M^l$ (the most rude estimation).
The rude estimation still includes
the cases where the total colors over the
pigeonholes gets less than $M-1$. 
We have to exclude the cases as, 
\begin{eqnarray*}
  F
  &=& M^l-[\text{total $(M-1)$ colors}]-[\text{total $(M-2)$ colors}] 
  \nonumber \\
  && - \cdots -[\text{total one color}] \ , 
\label{F_factor01}
\end{eqnarray*}
so that the total number should be just $M$. 
Noticing that 
[total ($M-k$) colors] under 
the factor ${}_G{C_M}$ 
just corresponds to 
$a(l,M-k;M)$ 
[{\it i.e.}, even providing $M$ choices, 
$l$ pigeonholes only amount to ($M-k$) colors
in total], 
we get 
\begin{eqnarray*}
F  &=& M^l - \sum\limits_{i = 1}^{M - 1}
          {a\left( {l,i;M} \right)} \ , 
\label{F_factor02}
\end{eqnarray*}
leading a recurrence, 
\begin{eqnarray}
  a\left( {l,M;G} \right){ = _G}{C_M}
  \cdot \left[ {{M^l} - \sum\limits_{i = 1}^{M - 1}
      {a\left( {l,i;M} \right)} } \right] \ .
\label{a.direct} 
\end{eqnarray}
We shall analyze the recurrence
further by the asymptotic evaluations
as below.
To check the asymptotic evaluations,
we can come back to the reccurrence
relation above, generating
a series $a\left( {l,M;G} \right)$  
by increasing $M=1,2,\cdots$
in a straightforward manner.

\vspace{2mm}
The denominator of Eq.~(\ref{p_first}) is 
evaluated as, 
\begin{eqnarray*}
  \sum_{i=1}^{G}{a(l,i;G)} = {G^l} \ , 
\label{nomalization}
\end{eqnarray*}
since each term under the summation 
corresponds to the case with 
'just equals to $i$ colors', 
counting of which over $i=1\sim G$
amounts to the total cases 
of assigning $G$ colors into $l$ 
pigeonholes allowing any 
duplications ($=G^l$). 
$P\left( {l,M;G} \right)$ is therefore 
evaluated as 
\begin{eqnarray}
  P\left( {l,M;G} \right)
  = \frac{{_G{C_M} \cdot \left[ {{M^l} - \sum\limits_{i = 1}^{M - 1}
          {a\left( {l,i;M} \right)} } \right]}}{{{G^l}}} \ , 
  \label{prob_exact}
\end{eqnarray}
under the assumption. 

\subsection{Asymptotic evaluations}
\label{asymptotic}
We decompose the probability 
$P\left(l,M;G\right)$ ('just $M$ colors')
into 
\begin{eqnarray}
  P\left( {l,M;G} \right) = A\left( {l,M;G} \right)
  \cdot D\left( {l,M;G} \right) \ ,
\label{decomposition}
\end{eqnarray}
where 
$A\left( {l,M;G} \right)$ 
is the probability to get 
'more than $M$ colors ($\ge M$)', 
while $D\left( {l,M;G} \right)$ 
denotes the rate to get $M$ 
over the possible range of $(M\sim G)$ 
for $A$. 
$D(l,M;G)$ is easily described as 
\begin{eqnarray}
  D\left( {l,M;G} \right)
  = \frac{{a\left( {l,M;G} \right)}}
  {{\sum\limits_{i = M}^G {a\left( {l,i;G} \right)} }} \ .
  \label{d01}
\end{eqnarray}

\vspace{2mm}
Considering the complementary of 
$A\left( {l,M;G} \right)$ ($\ge M$), 
the $Q=(1-A)$ corresponds to 
the case with [$\le (M-1)$]. 
If we define 
\begin{eqnarray}
  Q\left( {l,M;G} \right)
  = \frac{{\sum\limits_{i = 1}^{M} {a\left( {l,i;G} \right)} }}
  {{{G^l}}} \ , 
  \label{Q01}
\end{eqnarray}
the complementary relation is expressed as 
\begin{eqnarray}
  A\left( {l,M;G} \right) &= 1 - Q\left( {l,M-1;G} \right) , 
\label{cospace}
\end{eqnarray}
where we intentionally shift the definition 
of $M$ for $Q$ so that the upper limit of 
the summation in Eq.~(\ref{Q01}) to be $M$. 

\vspace{2mm}
Considering $Q\left(l+1,M;G\right)$ 
based on Eq.~(\ref{Q01}), 
we get, 
\begin{eqnarray}
  \frac{\sum\limits_{i = 1}^M {a\left(l+1,i;G\right)}}
        {G^{l + 1}}
       = \frac{{G\sum\limits_{i = 1}^M {a\left( {l,i;G} \right)}
           - \left( {G - M} \right)a\left( {l,M;G} \right)}}
           {{{G^{l + 1}}}} , 
\label{l01}
\end{eqnarray}
which is understood as follows: 
Imagine that 
the already existing $l$-pigeonholes 
include total $M$ colors, being 
corresponding to 
the first summation appearing in 
the nominator in the right-hand side. 
Multiplying $G$ to the summation 
means the counting all the possibilities 
for $(l+1)$-pigeonhole to have any color. 
That includes the cases with total $(M+1)$ 
colors, those should be excluded 
since we are considering the cases 
with 'total $M$ colors' 
because of the argument of $M$ in 
$Q\left(l+1,M;G\right)$. 
The part to be excluded is the product 
of $a(l,M;G)$ (the already existing pigeonholes 
have $M$ colors) times $(G-M)$ [$l+1$ pigeonhole 
gets a new color which is not found in 
the already existing pigeonholes]. 
With these meanings, we get the nominator 
of Eq.~(\ref{l01}). 

\vspace{2mm}
From the first term in the nominator of Eq.~(\ref{l01}), 
we make up $Q\left( {l,M;G} \right)$ to get, 
\begin{eqnarray}
  Q\left( {l + 1,G,M} \right)
  = Q\left( {l,M;G} \right) - \left( {\frac{{G - M}}{G}} \right)
  \frac{{a\left( {l,M;G} \right)}}{{{G^l}}} \ ,
\label{Q02}
\end{eqnarray}
as a recursion of $Q\left( {l,M;G} \right)$. 
Assuming the form, 
\begin{eqnarray}
  \left( {\frac{{G - M}}{G}} \right)
  \frac{{a\left( {l,M;G} \right)}}{{{G^l}}}
  =   \alpha\left(l,M;G\right) \cdot Q\left( {l,M;G} \right)  \ , 
\label{alpha}
\end{eqnarray}
and that $l$ to be treated as if 
it is continuous variable, then 
we get a differential equation, 
\begin{eqnarray}
  \frac{{\partial Q\left( {l,M;G} \right)}}{{\partial l}}
  =  - \alpha \left( {l,M;G} \right)Q\left( {l,M;G} \right)
  \nonumber \\
  \therefore\quad
  Q\left( {l,M;G} \right)
  = C\exp \left[ { - \int_1^l {\alpha \left( {l',G,M} 
  \right)dl'} } \right]  \ .
  \label{partial}
\end{eqnarray}
From Eq.(\ref{alpha}), the integrant $\alpha$ 
is asymptotically evaluated as, 
\begin{eqnarray*}
  \alpha\left(l,M;G\right)
  &=& \left( {\frac{{G - M}}{G}} \right)
  \frac{{a\left( {l,M;G} \right)}}
       {{\sum\limits_{i = 1}^M {a\left( {l,i;G} \right)} }}
  \sim  \left( {\frac{{G - M}}{G}} \right) \ , 
       \label{alpha01}       
\end{eqnarray*}
since the summation in the denominator 
is almost dominated by its leading term 
$a\left( {l,M;G} \right) \sim {M^l}$. 
This leads to an asymptotic form of $A\left( {l,M;G} \right)$ 
to be 
\begin{eqnarray}
  A\left( {l,M;G} \right)
  &=& 1 - C\exp\left[ { - \left( {\frac{{G - (M-1)}}
          {G}} \right)l} \right]{\rm{ }} \ .
\label{A_final}
\end{eqnarray}

\vspace{2mm}
Similarly, we shall derive an asymptotic 
form of $D\left( {l,M;G} \right)$ via 
differential approximations, 
finally as given in Eq.~(\ref{der.d}) 
by considering $D\left( {l+1,M;G} \right)$. 
We remind that the nominator and the denominator 
of Eq.(\ref{d01}) mean 
[Denominator; Number of cases with ($\ge M$) colors] and 
[Nominator; Number of cases with ($= M$) color], 
respectively, where we are considering the total 
number of colors ($M$ or whatever) included 
over pigeonholes.
Then we consider a newly added 
pigeonhole [$(l+1)$-th] on 
the already existing ones [$l$ pigeonholes]. 
We have two possibilities for the new 
pigeonhole those, 
'(x) new one doesn't gets a new color 
than those in the already existing pigeonholes' 
or 
'(y) new one gets a new color'. 
For the already existing pigeonholes, 
we sort the cases into three as 
'(a) the pigeonholes are getting 
$[1\sim (M-2)]$ colors', 
'(b) getting $(M-1)$ colors', 
and 
'(c) getting $M$ colors'. 
Since we are considering the cases 
with $(\ge M)$ only, 
the case (a) can be excluded from 
the present consideration 
[because it gives at most $(M-1)$ color 
when (a)*(y) occurs]. 
For (b) and (c), we finally get 
the conclusion that 
the contributions to 
$D\left( {l+1,M;G} \right)$ 
are made as 'Denominator; [(c)+(b)*(y)]' 
and 'Nominator; [(b)*(y)+(c)*(x)]', 
as explained as follows. 
For the denominator ($\ge M$), 
(c) contributes to it regardless of 
(c)*(x) [$= M$]
or (c)*(y) [$= (M+1)$]. 
Its 'number of cases' is then 
\[
\text{Denominator[(c)]} 
= G\cdot\sum\limits_{i = M}^G {a\left( {l,i;G} \right)} \ , 
\]
where $G$ is the number of the color-choice 
for $(l+1)$-th pigeonhole. 
For (b), only (b)*(y) contributes with $M$ colors, as 
\[
\text{Denominator[(b)*(y)]} = 
     [G - \left( {M - 1} \right) ]\cdot
a\left( {l,M-1;G} \right) \ , 
\]
where $[{G -\left( {M - 1}\right)}]$ is the 
possible choice for $(l+1)$-th one 
to have a new color not appearing in 
$l$ pigeonholes. 
Total contributions for the denominator 
amount to 
\begin{eqnarray}
&& \text{Denominator[(c)+(b)*(y)]} 
\nonumber \\
 && = G\sum\limits_{i = M}^G {a\left( {l,i;G} \right)}
+[G - \left( {M - 1} \right)] \cdot 
a\left( {l,M-1;G} \right) \ .
\end{eqnarray}
Similar way of counting for 
the nominator (num. of colors = $M$) 
gives the result of 
contributions, 
'Nominator[(b)*(y)+(c)*(x)]', 
where 
\begin{eqnarray}
 && \text{Nominator[(b)*(y)+(c)*(x)]}
\nonumber \\
  &&=
[G - \left( {M - 1} \right) ]\cdot
a\left( {l,M-1;G} \right)
+ M\cdot {a\left( {l,M;G} \right)}
\end{eqnarray}
By technically separating the 
second term in the nominator as $M=G-(G-M)$, 
we get
\begin{eqnarray*}
  && \frac{{a\left( {l + 1,M;G} \right)}}{{\sum\limits_{i = M}^G
      {a\left( {l + 1,i;G} \right)} }}
  \nonumber\\
  &=& \frac{{Ga\left( {l,M;G} \right)}}{{G\sum\limits_{i = M}^G
     {a\left( {l,i,G} \right)} + \left\{ {G - \left( {M - 1} \right)}
     \right\}a\left( {l,M-1;G} \right)}}
  \nonumber\\
  &&+ \frac{{ - \left( {G - M} \right)a\left( {l,M;G} \right)
      + \left\{ {G - \left( {M - 1} \right)} \right\}
      a\left( {l,M-1;G} \right)}}{{G\sum\limits_{i = M}^G
      {a\left( {l,i;G} \right)} + \left\{ {G - \left( {M - 1} \right)}
      \right\}a\left( {l,M-1;G} \right)}}
  \nonumber\\
  &=& \frac{{a\left( {l,M;G} \right)}}{{\sum\limits_{i = M}^G
      {a\left( {l,i;G} \right)} + \cfrac{{\left\{ {G - \left( {M - 1} \right)}
          \right\}a\left( {l,M-1;G} \right)}}{G}}}
  \nonumber\\
  &&+ \frac{{ - \cfrac{{\left( {G - M} \right)}}{G}a\left( {l,M;G} \right)
      + \cfrac{{\left\{ {G - \left( {M - 1} \right)} \right\}
          a\left( {l,M-1;G} \right)}}{G}}}{{\sum\limits_{i = M}^G
      {a\left( {l,i;G} \right)}  + \cfrac{{\left\{ {G - \left( {M - 1} \right)}
          \right\}a\left( {l,M-1;G} \right)}}{G}}} \ .
\label{D02}
\end{eqnarray*}
Since $a\left( {l,M;G} \right)\sim M^l$, 
we can neglect terms with $a\left( {l,M-1;G} \right)$ 
to get 
\begin{align*}
  \frac{{a\left( {l + 1,M;G} \right)}}{{\sum\limits_{i = M}^G
      {a\left( {l + 1,i;G} \right)} }}
  = \frac{{a\left( {l,M;G} \right)}}
  {{\sum\limits_{i = M}^G {a\left( {l,i;G} \right)} }}
  - \frac{{\left( {G - M} \right)}}{G}\frac{{a\left( {l,M;G} \right)}}
  {{\sum\limits_{i = M}^G {a\left( {l,i;G} \right)} }} \ , 
\end{align*}
leading to 
\begin{align*}
  D\left( {l + 1,M;G} \right)
  = D\left( {l,M;G} \right) - \left( {\frac{{G - M}}{G}} \right)
  D\left( {l,M;G} \right) \ .
\end{align*}
Regarding $l$ being continuous, we get 
\begin{eqnarray}
  \frac{{\partial D\left( {l,M;G} \right)}}{{\partial l}}
  &=&  - \left( {\frac{{G - M}}{G}} \right)D\left( {l,M;G} \right)
\label{der.d}
  \nonumber \\
\therefore \quad
  D\left( {l,M;G} \right)
  &=& \exp \left[ { - \left( {\frac{{G - M}}{G}}
      \right)\left(l-M\right)} \right] \ ,
\label{D_final}
\end{eqnarray}
where the constant of integration is 
determined so that $D\left( {l,M=1;G} \right)=1$ 
can be satisfied.

\begin{figure}[htbp]
  \includegraphics[width=\hsize]{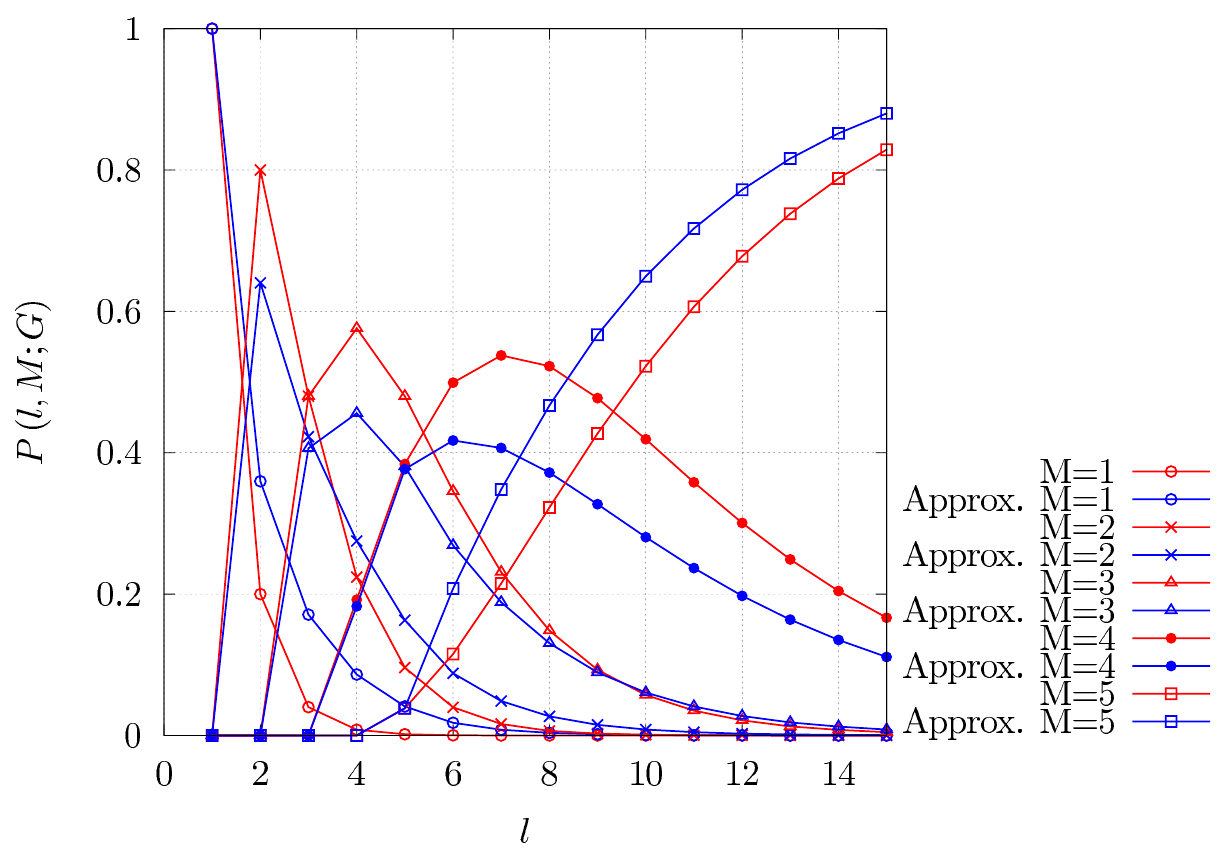}
  \caption{
    Comparisons between the asymptotic behavior [Approx., 
    Eq. (\ref{p30})] 
    and the original one [Eq. (\ref{p_first})] 
    of $P\left( {l,M;G} \right)$ 
    with $G=5$, as the functions of $l$ with 
    several different $M$. 
  }
  \label{P_approx}
\end{figure}

\vspace{2mm}
Substituting Eqs.(\ref{A_final}) and (\ref{D_final}) 
into Eq.(\ref{decomposition}), 
we get
\begin{align*}
  P\left( {l,M;G} \right)
  &\sim \left\{ {1 - C\exp \left[ { - \left( {\frac{{G - \left(M-1\right)}}
          {G}} \right)l} \right]{\rm{ }}} \right\}\\
  &\times \exp \left[ { - \left( {\frac{{G - M}}{G}} \right) 
      \left(l-M\right)} \right]
\end{align*}
as given in Eq.(\ref{prob_norm}) in the main text. 
At $l=M$, it reduces to 
\begin{align}
  P\left( {l,M;G} \right)_{l=M} = \frac{{_G{C_M} \cdot M!}}
  {{{G^M}}}  \ , 
  \label{p_rm01}
\end{align}
where it corresponds to the counting 
of possible permutations of $M$ chosen from $G$ 
colors, leading 
\begin{eqnarray}
    C = \left( {1 - \cfrac{{_G{C_M} \cdot M!}}{{{G^M}}}} \right) \ .
    \label{C01}
\end{eqnarray}
Since asymptotic evaluations have harmed  
the original normalization for $P$, 
we again introduce a normalization factor, 
\begin{align}
  Z\left( {l,G} \right)
  = \sum\limits_{M = 1}^G {P\left( {l,M;G} \right)} \ , 
\end{align}
by using of which 
\begin{eqnarray}
 P'\left( {l,M;G} \right)
  &=&\frac{1}{{Z\left( {l,G} \right)}}
  \left\{ {1 - C\exp \left[ { - \left( {\frac{{G - M + 1}}{G}} \right)
        \left( {l - M} \right)} \right]} \right\}
  \nonumber\\
  &&\times \exp \left[ { - \left( {\frac{{G - M}}{G}
        \left( {l - M} \right)} \right)} \right] \ , 
\label{p30}
\end{eqnarray}
becomes the probability to be applied to 
Eq.(\ref{aveAndvar}) in the asymptotic region. 
As shown in Fig.~\ref{P_approx}, 
the asymptotic evaluation performs 
fairly well to reproduce the behaviors 
of original form. 
Comparisons between the asymptotic behavior (Approx.) 
and the original one of $P\left( {l,M;G} \right)$ 
with $G=5$, as the functions of $l$ with 
several different $M$. 

\subsection{Identifying $l=l_0$ where the variance 
takes maximum}
\label{max.var}
By using Eq.(\ref{p30}) to evaluate the variance, 
Eq.(\ref{aveAndvar}), we get 
\begin{eqnarray}
  \frac{\partial }{{\partial l}}V\left[ M(l) \right]
  &=& \frac{1}{{{Z^2}}}\left[ {Z\sum\limits_{M = 1}^G
      {{M^2}\left( {\frac{{\partial P}}{{\partial l}}} \right)}
      - w\sum\limits_{M = 1}^G {{M^2}P}
      }\right.
      \nonumber\\
  && \left.{
      - 2\bar M\left\{ {Z\sum\limits_{M = 1}^G
        {M\left( {\frac{{\partial P}}{{\partial l}}} \right)}
        - w\sum\limits_{M = 1}^G {MP} } \right\}} \right] \ , 
  \label{var_diff2}
\end{eqnarray}
where we define 
\begin{align*}
  \bar M &= \sum\limits_{M = 1}^G {MP'}
  = \frac{1}{Z}\sum\limits_{M = 1}^G {MP}\\
  w &= \sum\limits_{M = 1}^G {\left( {\frac{{\partial P}}
      {{\partial l}}} \right)} \ , 
\end{align*}
and an approximation $C \approx 1$ is used. 
Since we can see as in the further appendix 
that $P$ and its derivative 
under the summations in each term of Eq.(\ref{var_diff2}) 
take steep peaks like $\sim \exp{(M-M_0)^2}$, 
we can evaluate them as 
\begin{eqnarray}
  F_2 &=& \sum\limits_{M = 1}^G {F(M)\cdot
    \frac{\partial P}{\partial l}}
  \sim {F(M_0^{(2)})\cdot
    \frac{\partial P}{\partial l}_{M=M_0^{(2)}}} \ ,
  \\
  F_1 &=& \sum\limits_{M = 1}^G {F(M)\cdot P\left( {l,M;G} \right)}
  \sim {F(M_0^{(1)})\cdot P\left( {l,M;G} \right)} \ ,
  \\
  Z   &=& \sum\limits_{M = 1}^G {P\left( {l,M;G} \right)}
  \sim {P\left( {l,M;G_0^{(0)}} \right)} \ .
\end{eqnarray}
For $F_2$, there are two peaks ($M_0 = G$ and $l$) 
as shown in \S\ref{der.f_dp01}, 
leading 
\begin{align}
  \sum\limits_{M = 1}^G {F\left( M \right)
    \left( {\frac{{\partial P}}{{\partial l}}} \right)}
  \sim
  F(G)\left( {\frac{{\partial P}}{{\partial l}}} \right)_{M=G}
  +
  F(l)\left( {\frac{{\partial P}}{{\partial l}}} \right)_{M=l} \ .
  \label{f_dp01}
\end{align}
By using an asymptotic evaluation of $\partial P/\partial l$ 
given in Eq.(\ref{prob_diff3}), 
we get 
\begin{align}
  \sum\limits_{M = 1}^G {F\left( M \right)
    \left( {\frac{{\partial P}}{{\partial l}}} \right)}
  = \frac{1}{G}\left[ {F\left( G \right)
      + \left\{ {G - l + 1} \right\}f\left( l \right)} \right] \ .
\label{f_dp03}
\end{align}
For $F_1$ and $Z$, 
the derivations given in \S\ref{der.f_dp01}) 
lead to 
\begin{eqnarray}
  \sum\limits_{M = 1}^G {F\left( M \right)P(L,M;G)}
  &\sim & F(l-1)\cdot P(L,M;G)_{M=(l-1)} \ ,
  \nonumber \\
  \sum\limits_{M = 1}^G {P(L,M;G)}
  &\sim & P(L,M;G)_{M=(l-1)}   \ . 
  \label{p01}
\end{eqnarray}
By defining 
\[P{\left( {l,M;G} \right)_{M = \left( {l - 1} \right)}}
\approx \left( {1 - {e^{ - \frac{{\left( {G - l + 1} \right)}}{G}}}}
\right){e^{ - \frac{{\left( {G - l + 1} \right)}}{G}}}
= :g\left( l \right) \ , \]
each term in Eq.(\ref{var_diff2}) is evaluated as 
\begin{eqnarray}
  && Z \sim  g\left(l\right)
  \quad , \quad 
  \bar M \sim  (l - 1)
  \quad , \quad   
  w \sim  \frac{{G - l + 2}}{G} \ ,
\nonumber \\
  &&\sum\limits_{M = 1}^G {MP} \sim \left( {l - 1} \right) g\left(l\right)
  \quad , \quad    
  \sum\limits_{M = 1}^G {{M^2}P} \sim {\left( {l - 1} \right)^2}
  g\left(l\right) \ , 
\nonumber \\
 && \sum\limits_{M = 1}^G {M\left( {\frac{{\partial P}}{{\partial l}}} \right)}
\sim \cfrac{1}{G}\left[ {G + \left( {G - l + 1} \right)l} \right] \ ,
\nonumber \\
 && \sum\limits_{M = 1}^G {{M^2}\left( {\frac{{\partial P}}{{\partial l}}}
    \right)} \sim \cfrac{1}{G}\left[ {{G^2} + \left( {G - l + 1}
      \right){l^2}} \right] ' 
\end{eqnarray}
respectively. 
Substituting them into Eq.(\ref{var_diff2}) 
leads to 
\begin{align*}
  \frac{\partial V}{{\partial l}}
  & =\frac{1}{{g\left( l \right)}}
  \left[ {\frac{ {{G^2}
        + \left( {G - l + 1} \right){l^2}} }{G}
      - \frac{{\left( {G - l + 2} \right)}}{G}{{\left( {l - 1} \right)}^2}}
    \right.\\
    & ~ \left. { - 2\left( {l - 1} \right)\left\{
      {\frac{ {G + \left( {G - l + 1} \right)l} }{G}
        - \frac{{\left( {G - l + 2} \right)}}{G}\left( {l - 1} \right)}
      \right\}} \right]\\
  &= \frac{1}{{Gg}}\left\{ {{l^2} - \left( {3 + 2G} \right)l
    + \left( {2 + 3G + {G^2}} \right)} \right\}
  \\
  & \propto [l-(G+1)][l-(G+2)] \ , 
\end{align*}
to get a conclusion that 
the variance has the maximum at 
$l=(G+1)$ or $(G+2)$. 

\subsection{Peaks of $P$ and its derivative}
\label{der.f_dp01}
Taking the the derivative of Eq.(\ref{prob_norm}), 
the approximation, $({G - M + 1}) \sim ({G - M})$, 
leads to 
\begin{align}
  &\frac{\partial }{{\partial l}}P\left( {l,M;G} \right)
  \nonumber\\
  &\sim \frac{1}{G}{e^{ - \frac{{\left( {G - M} \right)
  \left( {l - M} \right)}}{G}}}
  \left[ {\left\{ {2\left( {G - M} \right) + 1} \right\}
      {e^{ - \frac{{\left( {G - M} \right)\left( {l - M} \right)}}{G}}}
      - \left( {G - M} \right)} \right] \ .
  \label{prob_diff2}
\end{align}
The factor ${e^{ - \frac{{\left( {G - M}\right)\left({l - M} \right)}}{G}}}
\sim e^{-M^2}$ takes its maximum value $(=1)$ at the peak 
otherwise vanishes steeply. 
The value is realized when $M=G$ or $M=l$, 
accompanied by the value of the derivative, 
\begin{align}
  \frac{\partial }{{\partial l}}P\left( {l,M;G} \right)
  \sim \frac{1}{G}
  \left[ \left( {G - M} \right) + 1\right] \ , 
  \label{prob_diff3}
\end{align}
to be picked up from the summation 
as the product with the $F(M)$ in Eq.(\ref{f_dp01}). 

\vspace{2mm}
With $({G - M + 1}) \sim ({G - M})$ again 
applied to Eq.(\ref{prob_norm}), it leads to 
\begin{align*}
  P\left( {l,M;G} \right) \sim
  \left( {1 - {e^{ - \frac{{\left( {G - M} \right)
  \left( {l - M} \right)}}{G}}}}
  \right){e^{ - \frac{{\left( {G - M} \right)
  \left( {l - M} \right)}}{G}}} \ .
\label{p05}
\end{align*}
Taking $\partial P/\partial M = 0$ to get 
the condition for the peak, we obtain 
\begin{eqnarray*}
  \left( {G - M} \right)\left( {l - M} \right)
  &=& G\cdot\ln 2
\nonumber \\
\therefore 
    {M^2} - \left( {l + G} \right)M
    + G\left( {l - \ln 2} \right) &=& 0
\end{eqnarray*}
To simplify the factorization, 
we take further approximations 
on the term $G\left( {l - \ln 2} \right)$ 
as $G\sim (G+1)$ and $\ln 2\sim 1$, 
getting 
\[{M^2} - \left( {l + G} \right)M + \left( {G + 1} \right)
\left( {l - 1} \right) = 0 \ , \]
which is easily factorized as
\[\left[ {M - \left( {G + 1} \right)} \right]
\left[ {M - \left( {l - 1} \right)} \right] = 0 \ . \]
This provides a conclusion that 
the peak occurs at $M\approx (l-1)$, 
leading to Eq.(\ref{p01}) 
[another root, $M=G+1$, does not match with 
the setting of the problem, $M<G$].

\bibliographystyle{apsrev4-1}
\bibliography{references}
\end{document}